\documentclass[%
reprint,
nobibnotes,
 amsmath,amssymb,
 prl,
floatfix,
]{revtex4-1}

\usepackage{xcolor}
\usepackage{graphicx}
\usepackage{dcolumn}
\usepackage{bm}
\usepackage[mathlines]{lineno}

\usepackage{bm}
\newcommand{\tavg}[1]{\left\langle #1 \right\rangle}
\newcommand{\davg}[1]{\left\langle #1 \right\rangle_{\bm{J}}}

\newcommand{\sech}{\text{sech}}
\newcommand{\ca}{{\sim}}

\begin{document}

\title{Dimension of Activity in Random Neural Networks}
\author{David G. Clark}
\email{dgc2138@cumc.columbia.edu, he/him}
\author{L.F. Abbott}
\email{lfa2103@columbia.edu}
\author{Ashok Litwin-Kumar}
\email{a.litwin-kumar@columbia.edu}
\affiliation{Zuckerman Institute, Department of Neuroscience, Columbia University, New York, New York 10027, USA}
\date{\today}

\begin{abstract}
Neural networks are high-dimensional nonlinear dynamical systems that process information through the coordinated activity of many connected units. Understanding how biological and machine-learning networks function and learn requires knowledge of the structure of this coordinated activity, information contained, for example, in cross covariances between units. Self-consistent dynamical mean field theory (DMFT) has elucidated several features of random neural networks---in particular, that they can generate chaotic activity---however, a calculation of cross covariances using this approach has not been provided. Here, we calculate cross covariances self-consistently via a two-site cavity DMFT. We use this theory to probe spatiotemporal features of activity coordination in a classic random-network model with independent and identically distributed (i.i.d.) couplings, showing an extensive but fractionally low effective dimension of activity and a long population-level timescale. Our formulae apply to a wide range of single-unit dynamics and generalize to non-i.i.d. couplings. As an example of the latter, we analyze the case of partially symmetric couplings.

\end{abstract}

\maketitle

Neural circuits drive behavior, sensation, and cognition through the coordinated activity of many synaptically coupled neurons. Similarly, artificial neural networks solve tasks through distributed computations among neuronlike units with trained couplings. Understanding the structure of collective activity in such high-dimensional dynamical systems is a key problem in neuroscience and machine learning, made complicated by nonlinear units and heterogeneous couplings.

In addressing this problem, studying networks with random couplings has been fruitful \cite{sompolinsky1988chaos, stern2014dynamics, bahri2020statistical, poole2016exponential, huang2018mechanisms, PhysRevLett.129.048103}. Gradient descent dynamics often depend sensitively on the random initial couplings \cite{le2015simple, schoenholz2016deep, martens2021rapid, roberts2021principles}. Random couplings can provide a substrate for computation (see reservoir computing, \cite{jaeger2004harnessing, maass2007computational, rivkind2017local, susman2021quality}) and are a parsimonious model of background connectivity upon which structure can be introduced \cite{mastrogiuseppe2018linking}. In certain cases, trained networks learn low-rank additions to random connectivity \cite{schuessler2020interplay, martin2021implicit}. Finally, chaotic random networks model asynchronous cortical dynamics observed \textit{in-vivo} \cite{arieli1996dynamics}. More broadly, high-dimensional nonlinear dynamical systems with quenched disorder are important models in physics and ecology \cite{sompolinsky1988chaos, flyvbjerg1993mean, ciuchi1996self, aoki2014nonequilibrium, kadmon2015transition, aljadeff2015transition, chen2018dynamical, mastrogiuseppe2018linking, roy2019numerical, mignacco2020dynamical, keup2021transient, van2021large, krishnamurthy2022theory, de2022dynamical}.

Such disordered dynamical systems are commonly studied using dynamical mean-field theory (DMFT), which reduces the dynamics to a single-site problem, allowing for self-consistent calculation of single-unit temporal statistics. However, key properties of the structure of collective activity are not visible in single units, but only in their correlations. One such property is the effective dimension, which measures the degree of coordination of network activity via the approximate number of excited collective modes \cite{rajan2010inferring, abbott2011interactions, huang2018mechanisms, recanatesi2019dimensionality, engelken2020lyapunov, recanatesi2020scale, hu2022spectrum, jazayeri2021interpreting}. This quantity determines a network's ability to classify inputs \cite{litwin2017optimal, chung2018classification, cohen2020separability, farrell2022gradient}, learn via Hebbian plasticity \cite{litwin2017optimal, sorscher2021geometry}, generalize learned structure \cite{litwin2017optimal, sorscher2021geometry, cohen2022soft}, and generate dynamics \cite{sussillo2009generating, susman2021quality}. If the effective dimension is low, the network state can be inferred from a small number of units \cite{ganguli2012compressed, gao2015simplicity, gao2017theory, trautmann2019accurate}.

{Correlations and dimensionality have been studied in random feedforward networks \cite{poole2016exponential,huang2018mechanisms}, but such approaches do not generalize to recurrent dynamics due to the need to enforce self-consistency of network activity.} In this Letter, we develop a two-site DMFT, based on the cavity method, for high-dimensional nonlinear dynamical systems with quenched disorder, yielding a mean-field picture of a perturbatively coupled pair of units through which joint statistics are determined. The calculation applies across a broad range of dynamics for the individual units; we assume only that single-unit order parameters can be computed through usual DMFT techniques. Although our results are thus quite general, we apply them to the network model of \citet{sompolinsky1988chaos} with independent and identically distributed (i.i.d.) couplings, which displays a generic transition to chaos at a critical coupling variance \cite{kadmon2015transition}. We show analytically that collective activity is predominantly confined to a subspace of extensive but fractionally low dimension, previously observed only in simulations \cite{rajan2010inferring, abbott2011interactions, engelken2020lyapunov}. Our theory also reveals that collective modes have a typical timescale much longer than that of individual units. Finally, we show that our theory can capture the effect of non-i.i.d. connectivity that could arise through learning and analyze the case of partially symmetric couplings. 
 
\textit{Model \& Order Parameters:} We study a network of $N$ units with pre-activations $x_i(t)$ and activations ${\phi_i(t) = \phi(x_i(t))}$, where $\phi(\cdot)$ is a nonlinearity. The network has quenched disorder in its couplings, ${J_{ij} \overset{\text{i.i.d.}}{\sim} \mathcal{N}(0, g^2/N)}$. The network dynamics are
\begin{equation}
T[x_i](t) = \eta_i(t), \:\:\: \eta_i(t) = {\sum_j} J_{ij}\phi_j(t),
\label{eq:networkdynamics}
\end{equation}
where $T[\cdot]$ is a causal functional that specifies the single-unit dynamics, allowing for generalization beyond the conventional case of \cite{sompolinsky1988chaos}, ${T[x](t) = (1 + \partial_t) x(t)}$, to models with complex single-unit dynamics (e.g., \cite{stern2014dynamics,  marti2018correlations, muscinelli2019single, beiran2019contrasting, krishnamurthy2022theory, clark2023theory}). Our calculation applies to both linear and nonlinear $T[\cdot]$. In the latter case, individual pre-activations $x_i(t)$ may be highly non-Gaussian. We assume the system is temporally fluctuating and statistically stationary.

The classic DMFT of \cite{sompolinsky1988chaos} calculates the self-averaging autocovariance (two-point) function ${C^{\phi}(\tau) = \tavg{\phi_i(t) \phi_i(t + \tau)}_t = \davg{{\phi_i(t) \phi_i(t + \tau)}}\hspace{-1em}}$ (similarly, $C^{x}(\tau)$) through an effective single-site picture. For a given realization of $\bm{J}$, to zeroth order in $1/\sqrt{N}$, the individual local fields $\eta_i(t)$ are Gaussian with vanishing cross covariances. The network thus decouples into $N$ noninteracting processes, $T[x](t) = \eta(t)$, where $\eta(t)$ is a Gaussian field with zero mean and autocovariance $C^\eta(\tau) = g^2 C^\phi(\tau)$. The problem is closed by self-consistently enforcing $C^{\phi}(\tau) = \tavg{\phi(t) \phi(t + \tau)}_{\eta}$. $C^x(\tau)$ may be determined subsequently. In the model of \cite{sompolinsky1988chaos}, this single-site problem can be solved analytically due to the linearity of $T[\cdot]$ (in general, it is amenable to numerical solution; see, e.g., \cite{stern2014dynamics, roy2019numerical, mignacco2020dynamical, krishnamurthy2022theory, clark2023theory}).

In this Letter, we go beyond this single-site picture by calculating the structure of time-lagged cross covariances between units, $C_{ij}^\phi(\tau) = \tavg{\phi_i(t) \phi_j(t + \tau)}_t \sim 1/\sqrt{N}$ (similarly, $C_{ij}^x(\tau)$), ${i \neq j}$. These are non-self-averaging, so we examine the self-averaging four-point function
\begin{eqnarray}
\psi^\phi(\bm{\tau}) = N\davg{C^\phi_{ij}(\tau_1) C^\phi_{ij}(\tau_2)}, \:\:\: i \neq j
\label{eq:psi_defn}
\end{eqnarray}
and, analogously, $\psi^x(\bm{\tau})$. Our main result is that, for $N \rightarrow \infty$, $\psi^\phi(\bm{\tau})$ is given in Fourier space by
\begin{align}
\label{eq:psi_phi}
    \psi^{\phi}(\bm{\omega}) &= \left(M(\bm{\omega}) - 1\right)C^{\phi}(\omega_1)C^{\phi}(\omega_2) , \\
    \text{where } M(\bm{\omega}) &= \frac{1}{\left| 1 - 2\pi g^2 S^{\phi}(\omega_1) S^{\phi}(\omega_2)\right|^2} \nonumber,
\end{align}
and ${S^{\phi}(\tau) = \tavg{\delta \phi_i(t) / \delta \eta_i(t-\tau)}_t = \davg{\delta \phi_i(t) / \delta \eta_i(t-\tau)}}$ is the self-averaging linear response function given, as in the single-site problem, by ${S^{\phi}(\omega) = C^{\eta \phi}(\omega) / C^{\eta}(\omega)}$ where $C^{\eta \phi}(\tau) = \tavg{\eta(t) \phi(t + \tau)}_{\eta}$ (by the Furutsu-Novikov theorem; Appendix A). For the $x$ variables, we show
\begin{align}
\label{eq:psi_x}
    \psi^{x}(\bm{\omega}) &= |U(\bm{\omega})|^2 C^{\phi}(\omega_1) C^{\phi}(\omega_2) \nonumber \\
    & + U( \bm{\omega} ) C^{x\phi}(\omega_1)C^{x\phi}(\omega_2) + \text{H.c.}, \\
    \text{where} \:\: U(\bm{\omega}) &= \frac{ 2\pi g^2 S^{x}(\omega_1)S^{x}(\omega_2)}{1 - 2\pi g^2 S^{\phi}(\omega_1) S^{\phi}(\omega_2) }. \nonumber
\end{align}
Here, $C^{x \phi}(\tau) = \tavg{x(t) \phi(t + \tau)}_{\eta}$ and $S^{x}(\tau)$ is defined in analogy with $S^{\phi}(\tau)$. Notably, once the two-point and linear response functions have been computed in the single-site picture (analytically or numerically), the four-point functions are given analytically by Eqs.~\ref{eq:psi_phi}~and~\ref{eq:psi_x}, which apply for all $T[\cdot]$. These new order parameters encode important aspects of collective activity.

\textit{Two-site cavity DMFT.}—\begin{figure}
    \centering
    \includegraphics[width=0.6\columnwidth]{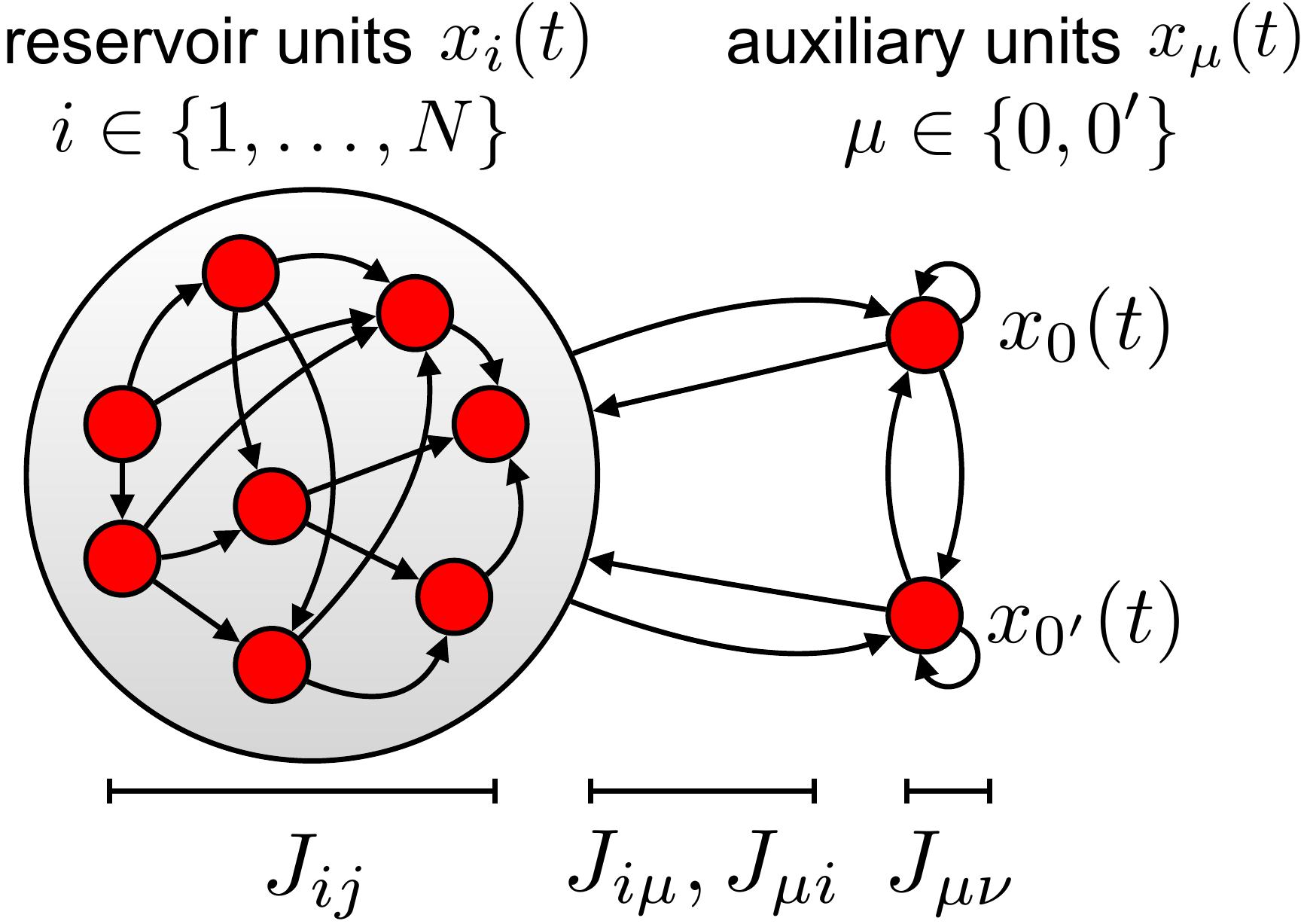}
    \caption{Overview of the two-site cavity DMFT. A cavity is created at the sites of two auxiliary units (called the cavity units) while the rest of the network, the reservoir, generates chaotic activity. The cavity units are then introduced; their effect is felt perturbatively by the reservoir. Dynamical equations for the cavity units yield a pair of perturbatively coupled mean-field equations representing a pair of units. Self-consistency conditions are constructed by noting that the cavity pair is statistically equivalent to any reservoir pair.}
    \label{fig:diagram}
\end{figure}Our derivation is based on a two-site \footnote{See \citet{mezard1987spin} Ch. V.3 for a two-site cavity approach to spin glasses.}, dynamical \cite{agoritsas2018out, roy2019numerical} version of the cavity method \cite{mezard1987spin, advani2013statistical} (Fig.~\ref{fig:diagram}).
We add two \textit{cavity units} to the network and refer to its original $N$ units as the \textit{reservoir}. We use ${i,j \in \{1, \dots ,N\}}$ and ${\mu,\nu \in \{0,0'\}}$ for reservoir- and cavity-unit indices, respectively. The cavity units are bidirectionally connected to the reservoir through $J_{i \mu}$ and $J_{\mu i}$, and to one another through\ $J_{\mu \nu}$. In the absence of the cavity units,
reservoir units follow trajectories $\phi_i(t)$. When the cavity units are introduced, these are perturbed by
$
    \delta \phi_i(t) = \int^t dt' \sum_j S^{\phi}_{ij}(t, t') \sum_{\mu} J_{j\mu}\phi_{\mu}(t'),
$
$S^{\phi}_{ij}(t, t') = {\delta \phi_i(t)}/{\delta \eta_{j}(t')}$. Inserting the perturbed trajectories ${\phi_i(t) + \delta \phi_i(t)}$ in Eq.~\ref{eq:networkdynamics} yields cavity-unit dynamics,
\begin{multline}
T[x_{\mu}](t) = \eta_{\mu}(t) + \frac{1}{\sqrt{N}}
\sum_{\nu} \int^t dt' F_{\mu \nu}(t, t')\phi_{\nu}(t'),
\label{eq:cavdynamics}
\end{multline}
where we have defined the order-one variables
\begin{align}
\eta_{\mu}(t) &= {\sum_i} J_{\mu i} \phi_i(t),
\label{eq:defn1} \\
\hspace{-.75em} F_{\mu \nu}(t, t') &= \sqrt{N}\Bigg[
{\sum_{ij}} J_{\mu i} J_{j\nu} S^\phi_{ij}(t, t') + J_{\mu \nu}\delta(t - t')\Bigg]. \label{eq:defn2}
\end{align}
In these definitions, the couplings and dynamic variables are mutually independent due to the cavity construction. Here, $\eta_{0}(t)$ and $\eta_{0'}(t)$ are \textit{cavity fields}, the local fields felt by the cavity units when they are not coupled to the reservoir. Due to the independence property, $\eta_{0}(t)$ and $\eta_{0'}(t)$ are jointly temporally Gaussian, with ${\sim 1/\sqrt{N}}$ cross covariance, to first order in $1/\sqrt{N}$ \footnote{By this, we mean that, for a given realization of $\bm{J}$, all (cross-)cumulants (under the $t$ average) higher than second-order are suppressed by at least $1/N$. Note that this joint Gaussianity property is \textit{not} true of the local fields $\eta_i(t)$ appearing in {Eq.~\ref{eq:networkdynamics}} due to correlations between $J_{ij}$ and $\phi_j(t)$.}.

Working in this two-site picture, we first determine an expression for
$C^{\phi}_{00'}(\omega)$. Using the joint Gaussianity of $\eta_0(t)$ and $\eta_{0'}(t)$ and the decoupling of sites $0$ and $0'$ under the time average, both valid to first order in $1/\sqrt{N}$, gives
\begin{align}
    &C^{\phi}_{00'}(\omega) = 2\pi \Big[ |S^{\phi}(\omega)|^2 C^{\eta}_{00'}(\omega) \nonumber \\
    &+ \frac{1}{\sqrt{N}} \left[ \left(F_{00'}(\omega)S^{\phi}(\omega)\right)^* + F_{0'0}(\omega)S^{\phi}(\omega) \right] C^{\phi}(\omega) \Big], \label{eq:tosquare} \\
    &\text{where } C^{\eta}_{00'}(\omega) = \sum_{ij} J_{0i} J_{0'j} C_{ij}^\phi(\omega), \label{eq:def3} 
\end{align}
and $F_{\mu \nu}(\tau) = \tavg{F_{\mu\nu}(t, t-\tau)}_t$ (Appendix B). We square and $\bm{J}$-average Eq.~\ref{eq:tosquare}, yielding $\psi^\phi(\bm{\omega})$. \hspace{-.5em}$F^*_{00'}(\omega)$, $F_{0'0}(\omega)$, and
$C^{\eta}_{00'}(\omega)$
are non-self-averaging; their six two-point functions under the $\bm{J}$ average are
\begin{subequations}
\label{eq:gamma_all}
\begin{align}
    \Gamma_{F_{00'} F_{00'}}(\bm{\omega}) &= \davg{F_{00'}(\omega_1) F_{00'}(\omega_2)}, \\
    \Gamma_{F^*_{00'} F_{0'0}}(\bm{\omega}) &= \davg{F^*_{00'}(\omega_1) F_{0'0}(\omega_2)}, \\
    \Gamma_{F^*_{00'} C^{\eta}_{00'}}(\bm{\omega}) &= \sqrt{N} \davg{F^*_{00'}(\omega_1) C^{\eta}_{00'}(\omega_2)}, \\
    \Gamma_{C^{\eta}_{00'} C^{\eta}_{00'}}(\bm{\omega}) &= N \davg{C^{\eta}_{00'}(\omega_1) C^{\eta}_{00'}(\omega_2)},
\end{align}
\end{subequations}
with the two others following from $0 \leftrightarrow 0'$ symmetry. The $\bm{J}$ averages can be evaluated due to the independence of the couplings and dynamic variables in Eqs.~\ref{eq:defn2} and \ref{eq:def3}. The $\Gamma_{\cdots}(\bm{\omega})$ functions are determined self-consistently by noting that the cavity pair is statistically equivalent to any reservoir pair, closing the equations. For i.i.d. $\bm{J}$, all of these vanish except $\Gamma_{C^{\eta}_{00'} C^{\eta}_{00'} }(\bm{\omega})$ and $\Gamma_{F_{00'} F_{00'}}(\bm{\omega})$. The former is given immediately by
\begin{equation}
    \hphantom{x} \Gamma_{C^{\eta}_{00'} C^{\eta}_{00'} }(\bm{\omega})=g^4 \left(C^{\phi}(\omega_1)C^{\phi}(\omega_2) +\psi^{\phi}(\bm{\omega})\right).
\end{equation}
The latter requires self-consistency. Doing the $\bm{J}$ average,
\begin{align}
\label{eq:tosolve}
    \Gamma_{F_{00'}  F_{00'}}&(\bm{\omega}) 
    = g^4 S^{\phi}(\omega_1) S^{\phi}(\omega_2) \nonumber \\
    &+ g^4 N \tavg{S^{\phi}_{00'}(\omega_1) S^{\phi}_{00'}(\omega_2)}_{\bm{J}} 
    +  \frac{g^2}{2\pi},
\end{align}
where ${S^\phi_{\mu\nu}(\tau)} = {\tavg{S^{\phi}_{\mu\nu}(t, t-\tau)}_t}$. Evaluating and substituting ${S^{\phi}_{00'}(\omega) = 2\pi N^{-\frac{1}{2}} S^{\phi}(\omega)^2 F_{00'}(\omega)}$ in Eq.~\ref{eq:tosolve} gives
\begin{equation}
    \Gamma_{F_{00'}F_{00'}}(\bm{\omega}) =  \frac{g^2}{2\pi\left(1 - 2\pi g^2 S^{\phi}(\omega_1) S^{\phi}(\omega_2)\right)}.
\end{equation}
Then, squaring and $\bm{J}$-averaging Eq.~\ref{eq:tosquare} gives
\begin{align}
    &\psi^{\phi}(\bm{\omega}) = 
    (2\pi)^2 \Big[ g^4 |S^{\phi}(\omega_1) S^{\phi}(\omega_2)|^2
    \Gamma_{C^{\eta}_{00'} C^{\eta}_{00'}}(\bm{\omega}) \nonumber \\ 
    &\hspace{-.75em}+\Gamma_{F_{00'} F_{00'}}(\bm{\omega}) S^{\phi}(\omega_1) S^{\phi}(\omega_2) C^{\phi}(\omega_1)C^{\phi}(\omega_2) 
    +\text{H.c.} \Big],
\end{align}
whose solution gives Eq.~\ref{eq:psi_phi}. Similar steps recover Eq.~\ref{eq:psi_x}. 

\textit{Effective Dimension.}—\begin{figure}
    \centering
    \includegraphics[width=\columnwidth]{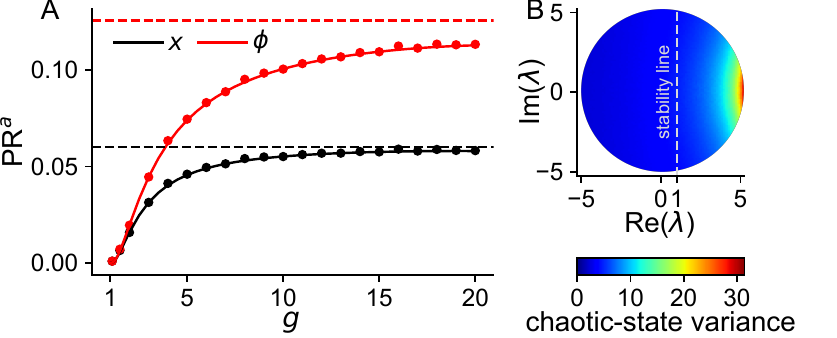}
    \caption{(a)
    Effective dimension $\text{PR}^a$ for $a \in \{x, \phi\}$. Lines: theory. Dots: simulations of size $N = 2500$. Median values over 50 disorder realizations are shown. (b) Activity variance in the chaotic state along eigenmodes of the stability matrix at the trivial fixed point with $g = 5$. Locally unstable modes with large real part have far greater variance in the chaotic state than those with small real part.} \label{fig:fig1}
\end{figure}We now specialize to the model of \cite{sompolinsky1988chaos}, with $T[x](t) = (1 + \partial_t) x(t)$ and $\phi(\cdot) = \tanh(\cdot)$, for which the single- and two-site pictures can be treated analytically (Appendix C). This model is chaotic for $g > 1$. We use Eqs.~\ref{eq:psi_phi} and \ref{eq:psi_x} to probe the structure of collective activity in the chaotic state, starting with the effective dimension. Let $\Sigma^a_{ij} = \tavg{a_i(t) a_j(t)}_t$ denote the covariance matrix of the variables $a \in \{x, \phi\}$ and $\lambda^a_i$ its eigenvalues. Following \cite{rajan2010inferring, abbott2011interactions, gao2015simplicity, gao2017theory, litwin2017optimal, recanatesi2019dimensionality, engelken2020lyapunov, recanatesi2020scale, hu2022spectrum} {(see \cite{huang2018mechanisms} for the feedforward-network case)}, we define the effective dimension as the participation ratio of this spectrum,
\begin{equation}
    \text{PR}^a = \frac{1}{N} \frac{ \left(\sum_i \lambda_i^a \right)^2}{ \sum_i (\lambda_i^a)^2},
\label{eq:prdef}
\end{equation}
where the factor $1/N$ makes this an intensive quantity, $0 \leq \text{PR}^a \leq 1$. $\text{PR}^a$ provides a linear notion of dimensionality, corresponding roughly to the minimal dimension, relative to $N$, of a subspace in which the strange attractor can be embedded with small L$_2$-norm distortion. This embedding subspace can be obtained from simulation results using principal components analysis (PCA) \cite{gao2015simplicity, gao2017theory, trautmann2019accurate}. We evaluate $\text{PR}^a$ for $N \rightarrow \infty$ by writing the numerator and denominator of Eq.~\ref{eq:prdef} as the squared trace and Frobenius norm, respectively, of $\bm{\Sigma}^a$, then expressing these quantities using DMFT order parameters, yielding
\begin{equation}
    \text{PR}^a = \frac{C^a(0)^2}{C^a(0)^2 + \psi^a(0, 0)}. 
\label{eq:prresult}
\end{equation}
Evaluating $\psi^a(0, 0)$ using Eqs.~\ref{eq:psi_phi}~and~\ref{eq:psi_x} yields agreement with simulations (Fig.~\ref{fig:fig1}A). $\text{PR}^a$ grows with $g$ as the activity becomes more tempestuous. $\text{PR}^{\phi} > \text{PR}^{x}$, reflecting the dimension-expanding effect of the nonlinearity, studied in feedforward networks \cite{babadi2014sparseness, litwin2017optimal,huang2018mechanisms}. Finally, $N\times\text{PR}^a \sim N$, consistent with ``extensive chaos'' \cite{engelken2020lyapunov}. 

In the ``Ising limit'' $g \rightarrow \infty$, $C^{\phi}(\tau)$, $\psi^{\phi}(\bm{\tau})$, $C^x(\tau)/g^2$, and $\psi^x(\bm{\tau})/g^4$ take on limiting forms. As a result, $\text{PR}^{\phi}$ and $\text{PR}^{x}$ saturate at finite values, $\text{PR}^x = 6.02\%$ and $\text{PR}^{\phi} = 12.6\%$ (Fig.~\ref{fig:fig1}A, dashed lines; Appendix C). Thus, the effective dimension is bounded substantially below one for arbitrarily large $g$, implying that structured couplings are required to increase the dimension further.

We next study $\text{PR}^a$ near the phase transition at $g = 1$, defining $\epsilon = g - 1 \ll 1$. To leading order in $\epsilon$, $C^a(0) = \epsilon$ and $\psi^a(0,0) = c/\epsilon$ where $c = F(0,0) = 4.27$ ($F$ described below). Thus, $\text{PR}^a = \epsilon^3 / c$. This scaling provides an interesting contrast with a linear stability analysis. Stability at the trivial fixed point, $x_i(t) = 0$, is determined by the spectrum of $\bm{J}$, which, at large $N$, is a uniform disk of radius $g$ \cite{girko1985circular, tao2008random}. The fractional area of the spectrum past the stability line, $\text{Re}(\lambda )= 1$, is $\ca\epsilon^{3/2}$. By contrast, $\text{PR}^a \sim \epsilon^3 \ll \epsilon^{3/2}$. Thus, locally unstable modes contribute to the chaotic state in a nonuniform manner. Simulations indicate that this nonuniform contribution of modes also holds at finite $\epsilon$ (Fig.~\ref{fig:fig1}B).

\textit{Temporal structure of cross covariances.}—\begin{figure}
    \centering
    \includegraphics[width=\columnwidth]{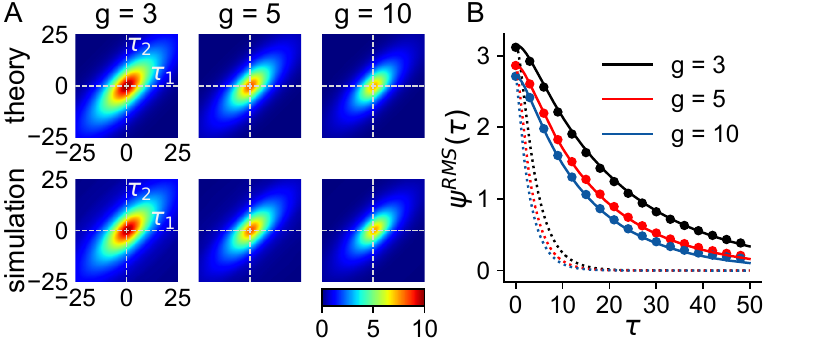}
\caption{
    (a) $\psi^{\phi}(\bm{\tau})$.
    (b) $\psi^{\text{RMS}}(\tau) = \sqrt{\psi^{\phi}(\tau, \tau)}$. Solid lines: theory. Dots: simulations. Dotted lines: rescaled single-unit autocovariance $C^{\phi}(\tau)$ for comparison.
    $N = 2500$. Median values across 50 disorder realizations are shown.
    } \label{fig:timescales}
\end{figure}We have focused on the effective dimension, expressed via $\psi^a(\bm{\tau})$ for $\bm{\tau} = (0,0)$. We now consider the full $\bm{\tau}$-dependence. Whereas $C^a(\tau)$ describes temporal structure of individual units, $\psi^a(\bm{\tau})$ describes temporal structure embedded in cross covariances. That $\psi^a (\tau, \tau) \neq \psi^a (\tau, -\tau)$ reflects the dissipative, time-irreversible nature of the network. The analytical form of $\psi^\phi (\bm{\tau})$, Eq.~\ref{eq:psi_phi}, agrees with simulations across $\bm{\tau}$ (Fig.~\ref{fig:timescales}A, B). Near the phase transition,
\begin{equation}
\psi^a(\bm{\tau}) = \frac{1}{\epsilon} F\left( \epsilon^2 \tau_+, \: \epsilon \tau_- \right), \:\: \tau_{\pm} = ({\tau_1 \pm \tau_2})/\sqrt{2},
\end{equation}
where $F(\bm{\tau})$ is a decaying order-one function (Appendix C). The timescale of $\psi^a(\bm{\tau})$ along the anti-diagonal, ${\tau_1  = -\tau_2}$, is $\ca1/\epsilon$, in agreement with the single-unit timescale \cite{sompolinsky1988chaos}. Strikingly, the timescale along the diagonal, ${\tau_1  = \tau_2}$, is $\ca1/\epsilon^2$, longer than the single-unit timescale by a (diverging) factor of $1/\epsilon$. This $1/\epsilon^2$ timescale is in agreement with the inverse Lyapunov exponent \cite{sompolinsky1988chaos}. This long timescale is also present at finite $\epsilon$: defining
$
{\psi^{\text{RMS}}(\tau) = \sqrt{\psi^{\phi}(\tau, \tau)}}
$,
we find that $\psi^{\text{RMS}}(\tau)$ has a much slower decay than $C^\phi(\tau)$ far from the transition (Fig.~\ref{fig:timescales}B).

As this long timescale is not present in individual units, it must arise in collective activity. This motivated us to examine, in simulations, the timescales of collective modes obtained by PCA. The timescales of PCs decrease across the variance-ranked PC index, implying that slow modes account for the most variance. As expected, the leading PCs have timescales many times longer than those of individual units (Fig.~\ref{fig:colltimescales}A). The density of timescales, $p(\tau)$, has an exponential tail. If this exponential form persists for $N \rightarrow \infty$, the timescales of leading PCs should diverge as $\ca\log N$. Simulating networks with sizes spanning two decades confirmed this (Fig.~\ref{fig:colltimescales}B). Thus, long timescales emerge in unstructured networks, albeit with only a $\log N$ divergence. Unlike in prior proposals (e.g., \cite{litwin2012slow, stern2014dynamics, marti2018correlations, berlemont2022glassy}), these long timescales are not visible in individual units, but arise at the collective level through the temporal structure of correlations.

\textit{Differential contributions to cross covariances.}—The cavity picture provides a partitioning of cross covariances into three sources, each contributing to $C_{00'}^\phi(\tau)$ at leading order, $1/\sqrt{N}$, in Eq.~\ref{eq:tosquare} (upon taking the inverse Fourier transform). First, units $0$ and $0'$ receive input from the same reservoir units, inducing a correlation between the cavity fields (Eq.~\ref{eq:tosquare}, $C^\eta_{00'}(\tau)$ term). Second, units $0$ and $0'$ have direct connections (Eq.~\ref{eq:tosquare}, delta-function terms in $F_{00'}(\tau)$ and $F_{0'0}(\tau)$). Third, unit $0'$ projects to the reservoir, producing reverberating activity read out by unit $0$, and vice versa (Eq.~\ref{eq:tosquare}, non-delta-function terms in $F_{00'}(\tau)$ and $F_{0'0}(\tau)$).
Isolating the terms in $\psi^\phi(\bm{\tau})$ corresponding to these three sources reveals that cavity-field correlations dominate near the phase transition, with the other two becoming larger further away from the transition (Fig.~\ref{fig:contributions}). Direct connections induce the shortest-timescale correlation. Reverberatory activity induces a slower correlation peaked at finite time lag. Cavity fields provide the slowest correlation. 

\textit{Structured disorder.}—Neural circuits undergo synaptic plasticity during learning. Thus, a key question is how structure in $\bm{J}$ shapes collective activity. Our calculation offers a natural method of incorporating structure in $\bm{J}$, namely, by enforcing it in the couplings involving cavity units ($0$ or $0'$) when self-consistently determining the two-point functions of Eq.~\ref{eq:gamma_all}. {This works when the structure in $\bm{J}$ is \textit{local} in the sense that the statistical structure of the entire $\bm{J}$ matrix is fully characterized by its effect on the couplings involving the cavity units.}

To demonstrate this, we calculated $\psi^\phi(\bm{\tau})$ under partially symmetric structure in the couplings, ${\tavg{J_{ij} J_{ji}} = g^2\rho / N}$, where $\rho$ is a symmetry parameter. In this case, $\psi^\phi(\bm{\tau})$ has the same form as Eq.~\ref{eq:psi_phi} but with
\begin{equation}
    M(\bm{\omega})
    \rightarrow M(\bm{\omega}) \frac{1 - \left|2\pi\sigma \left(S^\phi(\omega_1)\right)^* S^\phi(\omega_2) \right|^2}{\left| 1 - 2\pi\sigma \left(S^\phi(\omega_1)\right)^* S^\phi(\omega_2)\right|^2},
    \label{eq:psiphi_sym}
\end{equation}
where $\sigma = g^2 \rho$ (Appendix D). Additionally, rather than being negligible, the on-diagonal kernels $F_{00}(t, t')$ and $F_{0'0'}(t, t')$ are self-averaging with mean $\sqrt{N} \sigma S^\phi(t - t')$, resulting in an order-one self-coupling in the single-site problem. The equivalence of symmetric structure and an effective self-coupling is likely generic (e.g., such a term arises in networks with ongoing Hebbian plasticity but with the two-point function, rather than the linear response function, serving as the self-coupling kernel \cite{clark2023theory}).

Our theory yields agreement of $\text{PR}^{\phi}$ and $\psi^\phi(\bm{\tau})$ with simulations (Fig.~\ref{fig:symcase}). The effective dimension, $\text{PR}^\phi$, has a nontrivial relationship with the symmetry parameter $\rho$: near $\rho = 0$, increasing $\rho$ increases or decreases PR$^{\phi}$ depending on whether $g$ is small or large, respectively; for all $g$, making $\rho$ sufficiently large decreases PR$^{\phi}$.

Certain types of nonlocal structure can be handled with an expanded set of order parameters. One example is an intensive number groups of units with parameterized within- and across-group coupling statistics, modeling cell types in neural circuits (e.g., excitatory and inhibitory neurons). This could be generalized to a continuous group index, modeling spatial connectivity gradients. We find it unlikely that structure in $\bm{J}$ with nontrivial global constraints (e.g., orthogonality) could be handled by our cavity approach.

\textit{Discussion.}—We calculated the structure of time-lagged cross covariances in high-dimensional nonlinear dynamical systems with quenched disorder, allowing us to probe collective features of activity in chaotic neural networks.
Prior studies have analyzed cross covariances in noise-driven linear models with nonchaotic dynamics \cite{ginzburg1994theory, grytskyy2013unified, dahmen2019second, recanatesi2019dimensionality, yanliangspatial2022}. In this case, our theory readily recovers the frequency-dependent effective dimension ${\text{PR}^{x}(\omega) = (1 - 2\pi g^2 |S^x(\omega)|^2)^2}$ (see, e.g., Eqs.~10, 21 of \citet{hu2022spectrum}, who derived this from random matrix theory). While we used a cavity approach, deriving our results from fluctuations around the saddle point of a path integral would be interesting \cite{crisanti2018path, segadlo2021unified, grosvenor2022edge}.

Our calculation is agnostic about the single-unit dynamics $T[\cdot]$ and can account for structure in $\bm{J}$. It will be interesting to see how the dimension of activity is shaped by both types of structure. An important extension will be to incorporate time-dependent inputs, which can suppress chaos, an effect crucial to learning \cite{molgedey1992suppressing, sussillo2009generating, rajan2010stimulus, schuecker2018optimal}. In the case of inputs with low-dimensional structure, one expects a collapse from extensive to intensive effective dimension when chaos is suppressed. 

\begin{acknowledgements}
We thank Haim Sompolinsky for his advice on this work; Rainer Engelken, Samuel Muscinelli, and Sean Escola for comments; and James Murray and Sean Escola for early conversations. The authors were supported by the Gatsby Charitable Foundation and NSF NeuroNex award DBI-1707398. A. L.-K. was also supported by the McKnight Endowment Fund.
\end{acknowledgements}

\appendix
\renewcommand{\thefigure}{A.\arabic{figure}}
\setcounter{figure}{0}

\begin{figure}[ht]
    \centering
    \includegraphics[width=\columnwidth]{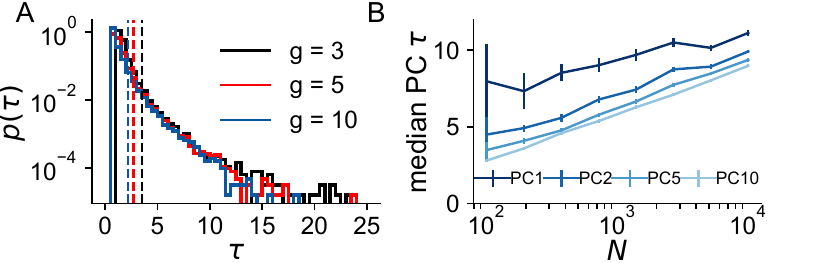}
\caption{
Timescales of $\phi$ principal components (PCs).
    (a)
    Density of PC timescales (autocovariance FWHM) in simulations. Dashed lines: single-unit timescales. $N = 2500$, 50 disorder realizations.
    (b)
    Timescales of leading PCs vs. $N$. $g = 10$. Median timescales across 50 disorder realizations are shown. Error bars: standard error of the mean.
    } \label{fig:colltimescales}
\end{figure}

\begin{figure*}[ht]
    \centering
    \includegraphics[width=\textwidth]{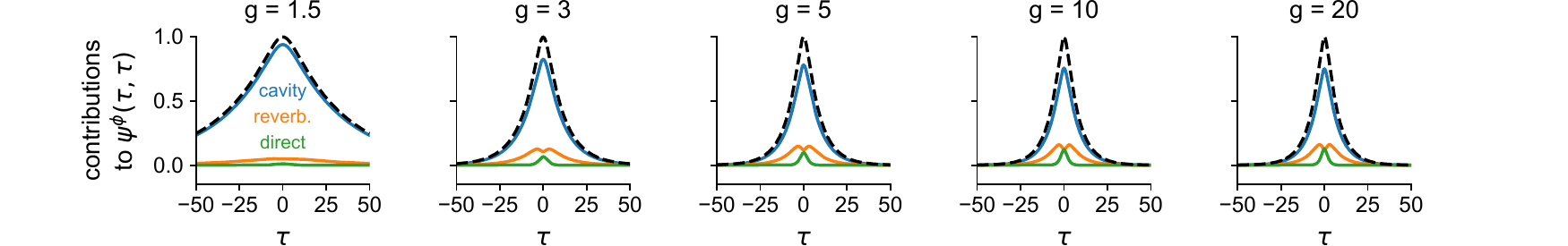}
    \caption{Relative contributions to $\psi^\phi(\tau, \tau)$ from distinct terms in the two-site cavity picture, namely, cavity field correlations (blue), reverberatory activity in the reservoir (orange), and direct coupling (green). Dashed line: $\psi^\phi(\tau, \tau)/\psi^\phi(0,0)$.}
    \label{fig:contributions}
\end{figure*}

\textit{Appendix A: Furutsu-Novikov theorem.}—Let ${T[x](t) = \eta(t)}$ where $\eta(t)$ is Gaussian with two-point function $C^{\eta}(t, t')$. Let $S^{\phi}(t, t') = \tavg{{\delta \phi(t)}/{\delta \eta(t') }}_{\eta}$, $\phi(t) = \phi(x(t))$. This can be expressed as a functional integral,
\begin{equation}
     S^{\phi}(t, t') = \int\mathcal{D} \eta e^{-S[\eta]}
    \frac{\delta \phi(t)}{\delta \eta(t') },
\end{equation}
where $S[\eta] = \frac{1}{2}\int ds \int ds' [C^{\eta}]^{-1}(s, s') \eta(s) \eta(s') +  \text{const}$. We write ${\mathcal{D} \eta =  d \eta({\neg t'}) d\eta(t')}$, where $d \eta({\neg t'})$ is the measure over all points on $\eta$ excluding $\eta(t')$. The integral over $\eta(t')$ can be evaluated via scalar integration by parts,
\begin{multline}
    \int d\eta(t') e^{-S[\eta]} \frac{\delta \phi(t)}{\delta \eta(t') }
    \\ = \int ds [C^{\eta}]^{-1}(t', s) \int d\eta(t') e^{-S[\eta]} \eta(s)\phi(t),
\end{multline}
where the boundary term vanishes for positive definite $C^{\eta}(t, t')$. 
Reintroducing the $d\eta(\neg t')$ integration gives
\begin{equation}
    S^{\phi}(t, t') = \int ds [C^{\eta}]^{-1}(t', s) C^{\eta \phi}(s, t),
\end{equation}
where $C^{\eta \phi}(s, t) = \tavg{\eta(s)\phi(t)}_{\eta}$. Assuming stationarity, we define $C^{\eta}(\tau) = C^{\eta}(t,t+\tau)$, $C^{\eta \phi}(\tau) = C^{\eta \phi}(t, t+\tau)$ and $S^{\phi}(\tau) = S^\phi(t, t-\tau)$. Then, $S^{\phi}(\tau) = ([C^{\eta}]^{-1} \ast C^{\eta\phi})(\tau)$. In Fourier space, $S^\phi(\omega) = C^{\eta\phi}(\omega)/C^{\eta}(\omega)$.

\textit{Appendix B: Deriving Eq.~\ref{eq:tosquare}.}—We first write a solution to Eq.~\ref{eq:cavdynamics} accurate to first-order in $1/\sqrt{N}$,
\begin{align}
\phi_{\mu}&(t) = \phi^{\text{free}}_{\mu}(t) \\
&+ \frac{1}{\sqrt{N}} \int^t dt' S^\phi_{\mu \mu}(t, t') \int^{t'}dt'' \sum_{\nu} F_{\mu \nu}(t', t'')\phi^\text{free}_{\nu}(t''),  \nonumber
\end{align}
where $T[x^{\text{free}}_{\mu}](t) = \eta_{\mu}(t)$ and $\phi^\text{free}_{\mu}(t) = \phi(x^\text{free}_{\mu}(t))$. We multiply the $0$ and $0'$ components and $t$-average, working to first order in $1/\sqrt{N}$. The two cross terms arising from $\phi^{\text{free}}_{\mu}(t)$ and the integral terms are readily evaluated by noting that the $0$ and $0'$ components decouple under the time average to zeroth order in $1/\sqrt{N}$, yielding the second and third terms in Eq.~\ref{eq:tosquare}. The nontrivial step is to evaluate $\tavg{\phi^\text{free}_0(t) \phi^{\text{free}}_{0'}(t+\tau)}_t$. Since $\eta_0(t)$ and $\eta_{0'}(t)$ can be treated as jointly temporally Gaussian with cross covariance $C^{\eta}_{00'}(\tau) \sim 1/\sqrt{N}$, we Taylor expand $\tavg{\phi^\text{free}_0(t) \phi^{\text{free}}_{0'}(t+\tau)}_t$ in $C^{\eta}_{00'}(\tau)$ via Price's theorem \cite{price1958useful}, which states (in a sufficiently abstract form) that, for $\bm{x} \sim \mathcal{N}(\bm{0}, \bm{\Sigma})$ and a function $F(\bm{x})$,
\begin{equation}
    \frac{\partial}{\partial \Sigma_{ij}} \tavg{F(\bm{x})}_{\bm{x}} = \tavg{\frac{\partial^2 F(\bm{x})}{\partial_{x_i} \partial_{x_j}} }_{\bm{x}},
\end{equation}
which is apparent upon writing the Gaussian integral corresponding to the LHS in Fourier space. Applying the functional version of this gives the first-order expansion
\begin{align}
    &\tavg{\phi^\text{free}_0(t) \phi^{\text{free}}_{0'}(t+\tau)}_t \nonumber \\ 
    &= \int ds \int ds' \tavg{\frac{\delta \phi^{\text{free}}_{0}(t)}{\delta \eta_0(s)} \frac{\delta \phi^{\text{free}}_{0'}(t+\tau)}{\delta \eta_{0'}(s')}}_t C_{00'}^{\eta}(s'-s) \nonumber \\
    &= \int ds \int ds' S^{\phi}(- s)S^{\phi}(\tau - s')C^{\eta}_{00'}(s' - s).
\end{align}
Denoting this by $C^{\phi^\text{free}}_{00'}(\tau)$, we have in Fourier space $C^{\phi^\text{free}}_{00'}(\omega) = 2\pi |S^{\phi}(\omega)|^2 C^{\eta}_{00'}(\omega)$, the first term in Eq.~\ref{eq:tosquare}. 

\textit{Appendix C: Sompolinsky et al. DMFT.}—The model of \cite{sompolinsky1988chaos} is recovered via $T[x](t) = (1 + \partial_t) x(t)$ and $\phi(\cdot) = \tanh(\cdot)$. Squaring the single-site picture, $(1 + \partial_t)x(t) = \eta(t)$, and averaging over $\eta$ reveals that $C^x(\tau)$ obeys Newtonian dynamics in a Mexican-hat potential, $V(C^x)$. $C^x(\tau)$ is obtained by numerically integrating this equation of motion. Then, $C^\phi(\tau)$ is given in Fourier space by ${C^\phi(\omega) = (1 + \omega^2) C^x(\omega) / g^2}$. The linear responses are $S^x(\omega) = 1/(1 + i\omega)$, $S^\phi(\omega) = \alpha / (1+i\omega)$ ($\alpha = \tavg{\phi'(x)}_{\eta}$).
Defining $X(\bm{\omega}) = (1 + i\omega_1)(1 + i\omega_2)$ and $\nu = g^2 \alpha^2$,
\begin{subequations}
\label{eq:specializdpsis}
\begin{align}
    \psi^\phi(\bm{\omega}) &= \left( \left| \frac{X(\bm{\omega})}{X(\bm{\omega}) - \nu} \right|^2 - 1 \right)C^\phi(\omega_1) C^\phi(\omega_2), \\
    \psi^x(\bm{\omega}) &= \left(  \frac{ 2|X(\bm{\omega})|^2 - \nu^2}{ \left| X(\bm{\omega}) - \nu \right|^2}  - 1 \right)C^x(\omega_1) C^x(\omega_2). \label{eq:psi_x_somp}
\end{align}
\end{subequations}
In evaluating Eq.~\ref{eq:psi_x} to obtain Eq.~\ref{eq:psi_x_somp}, we noted that ${C^{x \phi}(\omega) = \alpha C^x(\omega)}$ due to the Gaussianity of $x$. 

{Limit of ${g \rightarrow 1^+}$}:
Near the phase transition, $C^x(0) \sim \epsilon$ ($\epsilon = g- 1$). Taylor expanding $V(C^x)$ to fourth order in $C^x$ and analytically integrating the resulting equation of motion gives ${C^x(\tau) = C^\phi(\tau) = \epsilon \: \text{sech} \left( \tau \epsilon  / \sqrt{3} \right)}$ to first order in $\epsilon$ \cite{crisanti2018path}. The large-$\tau$ autocovariance behavior is controlled by the quadratic term in $V(C_x)$, so including higher-order terms in the Taylor expansion gives corrections decaying faster than $\ca1/\epsilon$. Thus, in Fourier space, ${C^a(\omega) = ({3 \pi/2})^{1/2}\text{sech}\left(\sqrt{3} \pi \omega / 2 \epsilon \right)}$ at leading order in $\epsilon$. Noting that ${\nu = 1 + V''(0)}$, we have, from the expanded potential, ${\nu = 1 - \epsilon^2 / 3}$ at leading nontrivial order. The divergent contributions to the inverse transforms of both $\psi^x(\bm{\omega})$ and $\psi^{\phi}(\bm{\omega})$ come from the term ${C^a(\omega_1)C^a(\omega_2)/|X(\bm{\omega})-\nu|^2}$. This contribution is
\begin{equation}
    \hphantom{a\hspace{-1em}} \psi^a(\bm{\tau}) = \frac{3 }{4} \int \hspace{-.35em} d\bm{\omega} e^{i \epsilon \bm{\omega}^T \bm{\tau}} \frac{\sech\left( \frac{\sqrt{3}\pi\omega_1}{2}  \right)\sech\left( \frac{\sqrt{3}\pi\omega_2}{2} \right)}{\epsilon^2 (1/3 \hspace{-.2em}-\hspace{-.2em} \omega_1\omega_2)^2 + (\omega_1 \hspace{-.2em}+\hspace{-.2em} \omega_2)^2}.
\end{equation}
Along the anti-diagonal, $\omega_1 + \omega_2 = 0$, the integrand diverges as $\ca1/\epsilon^2$. The divergent part is a ridge with thickness $\ca\epsilon$, corresponding to the first and second terms in the denominator having similar magnitudes. The $\ca1/\epsilon^2$ height and $\ca\epsilon$ thickness give a $\ca1/\epsilon$ divergence of $\psi^a(\bm{\tau})$. Defining $\omega_{\pm} = (\omega_1 \pm \omega_2)/\sqrt{2}$, we rotate the ridge onto the $\omega_-$ axis. The $\ca\epsilon$ thickness of the ridge along the $\omega_+$ axis suggests replacing $\omega_+ \rightarrow \epsilon \omega_+$, yielding
\begin{subequations}
\begin{align}
    \psi^a(\bm{\tau}) &= \frac{1}{\epsilon} \int d\bm{\omega}_{\pm} e^{i\left(\epsilon^2 \omega_+ \tau_+ + \epsilon \omega_- \tau_- \right)} \frac{F(\omega_+, \omega_-)}{2\pi}, \\
    F(\bm{\omega}_\pm) &= \frac{3\pi}{2} \frac{
    \text{sech}^2\left(\frac{\sqrt{3}\pi  \omega_-}{2^{3/2}} \right) }{\left(1/3 + \omega_-^2 / 2\right)^2 + 2 \omega_+^2},
\end{align}
\end{subequations}
where $\tau_\pm = ({\tau_1 \pm \tau_2})/{\sqrt{2}}$ as defined in the main text. 

{Limit of $g \rightarrow \infty$:}
Defining $\bar{C}^x(\tau) = C^x(\tau)/g^2$, the Newtonian dynamics in this limit approach ${\partial_{\tau}^2 \bar{C}^x(\tau) = \bar{C}^x(\tau) - ({2}/{\pi}) \arcsin \left({\bar{C}^x(\tau)}/{\bar{C}^x(0)} \right)}$ where $\bar{C}^x(0) = 2(1 - 2/\pi)$ \cite{crisanti2018path}. Meanwhile, $\nu = 1/(\pi - 2)$. Numerically integrating these dynamics allows us to evaluate Eq.~\ref{eq:specializdpsis} numerically, producing the saturating values of $\text{PR}^a$ given in the main text.

\begin{figure}[ht]
    \centering
    \includegraphics[width=\columnwidth]{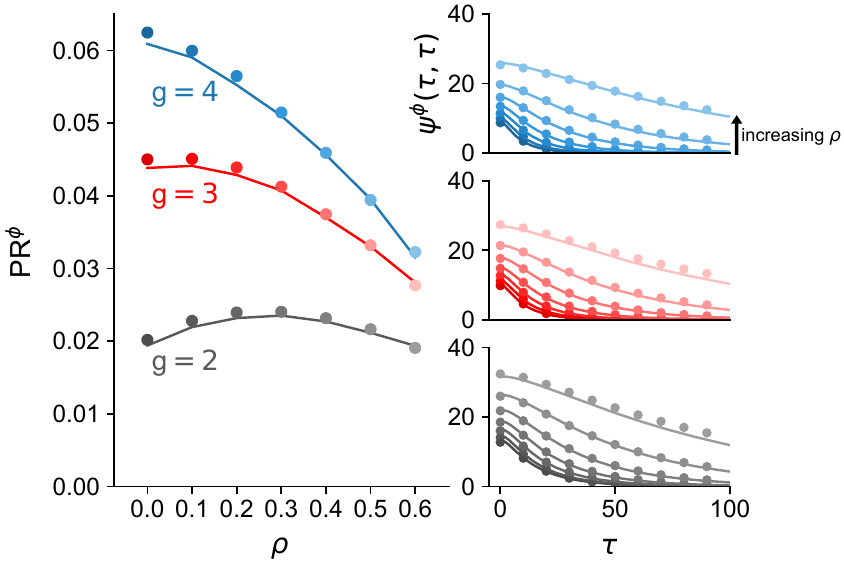}
    \caption{(a) Effective dimension PR$^{\phi}$ as a function of $\rho$ for various values of $g$. (b) $\psi^\phi(\tau, \tau)$ as a function of $\tau$ for values of $g$ and $\rho$ from (a). Lines: DMFT. Dots: simulations of size $N = 7500$, median across 200 realizations.
    }
    \label{fig:symcase}
\end{figure}

\textit{Appendix D: Partially symmetric disorder.}—For finite ${\sigma = g^2 \rho}$, $\davg{F_{\mu \mu}(t, t')} = \sqrt{N} \sigma S^\phi(t - t')$, modifying the single-site picture by introducing an order-one self-coupling. This self-coupling can be absorbed into the dynamics functional by replacing $T[\cdot]$ with
\begin{equation}
    T_{\sigma}[x](t) = T[x](t) + \sigma \int_{-\infty}^t dt' S^\phi(t-t') \phi(t').
\end{equation}
In our case, ${T[x](t) = (1 + \partial_t)x(t)}$. We solve the single-site DMFT numerically, using the Furutsu-Novikov theorem to evaluate $S^\phi(\tau)$.
We now self-consistently determine the two-point functions of Eq.~\ref{eq:gamma_all} under correlated disorder. $\Gamma_{F_{00'} F_{00'}}(\bm{\omega})$ and $\Gamma_{C^\eta_{00'} C^\eta_{00'}}(\bm{\omega})$ are the same as for of i.i.d. disorder. Rather than vanishing, we have also
\begin{subequations}
\begin{align}
   &\Gamma_{F^*_{00'} F_{0'0}}(\bm{\omega}) = \frac{\sigma}{2\pi\left(1 - 2\pi \sigma (S^\phi(\omega_1))^* S^\phi(\omega_2)\right)}, \\
   &\Gamma_{F^*_{00'} C^\eta_{00'} }(\bm{\omega}) \\
   &= (2\pi)^2 \Gamma_{F_{00'} F_{00'}}^*(\bm{\omega}) \Gamma_{F^*_{00'} F_{0'0}}(\bm{\omega}) (S^\phi(\omega_1))^* C^{\phi}(\omega_2). \nonumber 
\end{align}
\end{subequations}
Using these when squaring and $\bm{J}$-averaging Eq.~\ref{eq:tosquare} yields the solution for $\psi^\phi(\bm{\omega})$, Eq.~\ref{eq:psiphi_sym}. The DMFT predictions for $\text{PR}^\phi$ and $\psi^\phi(\bm{\tau})$ agree with simulations (Fig.~\ref{fig:symcase}). 

For $\rho$ approaching unity, this model \cite{marti2018correlations} and related models \cite{berlemont2022glassy} have been reported to exhibit temporally nonstationary, ``glassy'' dynamics, and it is possible that nonstationary behavior appears for $\rho < 1$. Our calculation assumes temporal stationarity, but nevertheless matches simulations up to $\rho = 0.6$ (modulo small discrepancies that are likely attributable to numerics). Numerically solving the single-site DMFT for larger values of $\rho$ is challenging due to the rapid growth of the autocorrelation timescale with $\rho$.

\bibliography{refs}

\end{document}